\documentstyle[12pt]{article}
\begin{document}
\title{  
\hfill \parbox{4cm}{\normalsize OPCT-94-3\\hep-lat/9412109}\\
\vspace{2cm}
        Low-Temperature Expansion \\ 
        of the Free Energy in ASOS Model 
       }
\author{
        H. ARISUE \\ 
        \normalsize\em
        Osaka Prefectural College of Technology \\ 
        \normalsize\em
        Saiwai-cho, Neyagawa, Osaka 572, Japan \\
        \normalsize\rm 
        Internet:arisue@ipc.osaka-pct.ac.jp
        }
\date{\normalsize December 1994}
\maketitle

\begin{abstract}
  We calculate the low-temperature series of the free energy 
in absolute-value solid-on-solid (ASOS) model 
to order $u^{23}$ using finite-lattice method.
 The property of the obtained series 
and the behavior of their Pad\'e approximants confirms us 
that the roughening transition in ASOS model is of 
Kosterlitz-Thouless type. 
\end{abstract}

\newpage
\section{Introduction}
 Recently the low-temperature series in $d=3$ Ising model 
or equivalently strong-coupling series in $d=3$ $Z_2$ 
lattice gauge theory have been extended to higher orders 
using finite-lattice method 
      \cite{Enting,Arisueone_two_six,Creutz}. 
 This method avoids the problem involved in the graphical method, 
in which it is rather difficult 
to give the algorithm for listing all the diagrams up completely 
that contribute to the relevant order of the series. 
 The low-temperature series calculated by finite-lattice method 
in $d=3$ Ising model are for the true inverse correlation length 
(which is equivalent to the mass gap in lattice gauge theory) 
      \cite{Arisuethree_five},
free energy 
      \cite{Bhanot,Guttmann},
magnetization and zero-field susceptibility
      \cite{Guttmann},
surface tension 
(which is equivalent to the string tension in lattice gauge theory) 
      \cite{Arisuefour}
and the second moment of the correlation function 
      \cite{Arisueseven}.
 We should mention that,  
using a modification of the shadow-lattice technique, 
low-temperature series were obtained 
for free energy, magnetization and susceptibility 
      \cite{Vohwinkel},
which are longer than those of reference~\cite{Bhanot,Guttmann}.
 The second moment was also calculated 
by the same technique 
      \cite{Vohwinkeltwo}, 
in which exact series is shorter and estimated series 
is longer than that of reference~\cite{Arisueseven}. 
 
 As for the surface tension we \cite{Arisuefour} calculated the 
low-temperature series to order $u_{_{Ising}}^{17}$ 
with $u_{_{Ising}}=\exp{(-4\beta_{_{Ising}})}$ 
and found that the series coefficient change their sign 
at the order of $u_{_{Ising}}^{13}$. 
 The surface tension in three-dimensional Ising model 
or the string tension in three- or higher-dimensional 
lattice gauge theory 
suffers from the roughening transition 
      \cite{Weekstwo,Hasenfratz,Itzykson} 
and it is expected to exhibit Kosterlitz-Thouless \cite{Kosterlitz}
type singularity like 
\begin{equation}
   f(u)=A(u) \exp{[-c(u_r-u)^{-1/2}]} + B(u)
      \label{KT}.
\end{equation}
 It has the essential singularity 
at the roughening transition point $u_r$. 
As was pointed out by Hasenbusch and Pinn 
      \cite{Hasenbuschtwo}, 
the sign-change of the coefficients is just the signal 
of the K-T type singularity. 
 In fact if we expand the function (\ref{KT}) in terms of $u$ 
we would obtain a series with sign-change. 
The order of the sign-change depends on $c$ and $u_r$. 
 The Pad\'e approximants of the low-temperature series 
for the surface tension appear to converge 
below the order of the sign-change \cite{Shaw}. 
 They, however, start to scatter just at the order of the sign-change 
and appears to converge again for higher orders \cite{Arisuefour}. 
 The Pad\'e approximants of the maximal-order series 
for the surface tension 
exhibit good convergence but their values overshoot the high precision 
Mote Carlo results 
      \cite{Hasenbuschtwo}. 
 We think that our series are yet too short for the asymptotic 
regime to be reached.
 Unfortunately it would be difficult to extend the series 
further by the presently available technique 
of the finite-lattice method and computer resources. 
 
 Absolute-value solid-on-solid (ASOS) model 
is an approximation of the interface of d=3 Ising model. 
 Overhangs and disconnected parts are neglected in the former. 
The ASOS model is also expected to exhibit K-T type phase transition. 
 Its free energy corresponds to the surface tension 
of $d=3$ Ising model and is expected to behave like~(\ref{KT}). 
 Hasenbusch and Pinn 
      \cite{Hasenbuschtwo} 
calculated the low-temperature series 
for the free energy of this model to order $u^{12}$ 
using the finite-lattice method, 
extending the previous series by Weeks et al 
     \cite{Weeks}, 
and they found the expected sign-change of the coefficients 
at the order of $u^{11}$. 
 They analyzed the series by Pad\'e approximants, 
compared with their Monte Carlo data 
and obtained the results similar to the case of the surface tension 
in $d=3$ Ising model. 
 They concluded that the Pad\'e approximants are inconsistent 
to the Monte Carlo data. 
 Only two terms after the sign-change, however, 
would be too short to make a definite conclusion. 

 In this paper we extend the low-temperature series 
for the free energy of ASOS model to order $u^{23}$ 
using the finite-lattice method. 
 An improved technique is introduced to extend the series, 
which can also be applied to the low-temperature expansion 
of the spin systems whose spin-variable takes more than 2 values 
such as the spin-$S$ Ising model with $S \ge 1$ 
or the $q$-state Potts model with $q \ge 3$.
 The obtained longer series enables us to make a definite conclusion 
on the roughening transition in the ASOS model 
and on the convergence of the Pad\'e approximants for the free energy. 
\section{Algorithm}
 Here we give the improved algorithm of the low-temperature expansion 
for SOS model that enables us to obtain longer series than the case 
when the finite-lattice method was applied naively.
 Let us consider two-dimensional 
$L_x \times L_y$ rectangular lattice $\Lambda_0$. 
The free energy density in the infinite-volume limit is given by 
\begin{equation}
    f =  
   - \lim_{\scriptstyle 
              L_x,L_y \rightarrow \infty,
           \atop\scriptstyle 
              H_{+},H_{-} \rightarrow \infty} 
          \frac{1}{ L_x L_y } 
         \ln{ \big[ Z(L_x,L_y;H_{+},H_{-}) \big] },
\end{equation}
with Hamiltonian
\begin{equation}
    Z(L_x,L_y;H_{+},H_{-})
       =  \sum_{\{h\}} 
             \exp{ \big( - \beta \sum_{<i,j>} | h_i - h_j | \big)  }
\end{equation}
where the variable $h_i$ at each site $i$ is 
restricted to $ -H_{-} \le h_i \le H_{+}$  $( H_{+},H_{-} > 0 )$. 
 The low-temperature series is calculated 
with respect to the expansion parameter $u=\exp{(-2\beta)}$. 
 We take the boundary condition 
that all the variables outside $\Lambda_0$ are fixed to be zero. 

 Let us consider the set $\{\Lambda\}$ 
of all two-dimensional rectangular sub-lattices of $\Lambda_0$. 
 The sub-lattice $\Lambda$ is denoted by its size $l_x \times l_y $ 
and its position in $\Lambda_0$.
We define $H$ of $\Lambda$ as 
\begin{equation}
   H(\Lambda;h_{+},h_{-}) 
     = - \ln{ \big[ Z(\Lambda;h_{+},h_{-}) \ \big] }     
      \label{defH}.
\end{equation}
In the calculation of the partition function 
$Z(\Lambda;h_{+},h_{-})$ 
the variable $h_i$ inside the $\Lambda$ is restricted 
to the range of $ -h_{-} \le h_i \le h_{+} \ ( h_{+}, h_{-} \ge 0 )$  
and all the variables outside are fixed to be zero. 
We define $W$ of $\Lambda$ recursively as
\begin{equation}
    W(\Lambda;h_{+},h_{-})
        = H(\Lambda;h_{+},h_{-}) 
               - {
         \sum_{\scriptstyle
                    \Lambda^{\prime} \subseteq \Lambda, 
               \atop\scriptstyle 
         h_{+}^{\prime} \le h_{+}, h_{-}^{\prime} \le h_{-} }
                  }^{ \hspace{-0.8cm} (')} \hspace{0.8cm}
           W( \Lambda^{\prime} ; h_{+}^{\prime},h_{-}^{\prime} ). 
\end{equation}
 Here the summation ${\sum}^{(')}$ implies that 
the $W(\Lambda;h_{+},h_{-})$ should be excluded in the summation.
 We note that $H(\Lambda;h_{+},h_{-})$ and $W(\Lambda;h_{+},h_{-})$ 
is independent of the position of $\Lambda$. 
 We know 
\begin{eqnarray}
   H(\Lambda_0;H_{+},H_{-}) 
    &=& \sum_{\scriptstyle
           \Lambda \subseteq \Lambda0,
               \atop\scriptstyle 
           \ h_{+} \le H_{+}, \ h_{-} \le H_{-}} 
                        W(\Lambda;h_{+},h_{-}) \nonumber \\
    &=& \sum_{\scriptstyle
                l_x \le L_x, \ l_y \le L_y, 
              \atop\scriptstyle 
                \ h_{+} \le H_{+}, \ h_{-} \le H_{-}}  
                         (L_x-l_x+1)(L_y-l_y+1)          \nonumber \\
    && \;\;\;\;\;\;\;\;\;\;\;\;\;\;\;\;\;\;\;\;\;\;\;\;\;\;\;\;\;
       \;\;\;\;\;\;\;\;\;\;\;\;\;
                \times W( l_x \times l_y;h_{+},h_{-} ).  
\end{eqnarray}
 Taking the infinite-volume limit we obtain 
\begin{eqnarray}
    f &=&  \lim_{ L_x,L_y \rightarrow \infty} 
            \frac{1}{ L_x L_y} H(\Lambda_0;H_{+},H_{-})    \nonumber \\
      &=&  \sum_{ l_x,l_y,h_{+},h_{-}} 
                  W( l_x \times l_y;h_{+},h_{-} ).                
      \label{free_energy}
\end{eqnarray}
 
 In the standard cluster expansion of the free energy for this model, 
a cluster is composed of polymers and  
each of the polymers consists of sites 
that are connected by nearest-neighbor bonds. 
 A value $h_i( \neq 0 )$ is put to each site $i$ of the polymer. 
 We can assign to each cluster two numbers $h_p$ and $h_m$ 
that are the maximum of $h_i(\ge 0)$ and $-h_i(\ge 0)$, respectively, 
in all the sites of the polymers the cluster consists of.
 Then, we can prove 
      \cite{Arisueone_two_six} 
that the Taylor expansion of $W(l_x \times l_y;h_{+},h_{-})$ 
with respect to $u$ includes 
the contribution from all the clusters of polymers 
in the standard cluster expansion 
that have $h_p=h_{+}$ and $h_m=h_{-}$ 
and that can be embedded into $l_x \times l_y$ lattice 
but cannot be embedded into any of its rectangular sub-lattices.
 The series expansion of $W(l_x \times l_y;h_{+},h_{-})$ 
starts from the order $u^n$ with 
$  n = l_x + l_y + 2(h_{+}+h_{-})-2 $. 
 So, to obtain the expansion series to order $u^N$,
we should take into account all the rectangular lattices 
that satisfy 
$ l_x + l_y + 2(h_{+}+h_{-})-2 \le N $
in the summation of eq.~(\ref{free_energy}) 
and evaluate each of the $W$'s to order $u^N$. 

 All the above algorithm can be coded on the computer 
and most of the CPU time and main memory are used 
for the evaluation of the relevant partition functions 
$Z(l_x \times l_y;h_{+},h_{-})$ to order $u^N$.
 The partition functions are calculated 
using the transfer matrix formalism based on building up finite-size 
lattices one site at a time 
      \cite{Entingtwo}.
 The necessary CPU time and memory are proportional 
to $ N l_x l_y \ (h_{+}+h_{-}+1)^{\min{(l_x,l_y)}} $
and $ N \ (h_{+}+h_{-}+1)^{\min{(l_x,l_y)}}$,
respectively. 
 In table~\ref{tab:memory} we list the maximum $M_{imp}(N)$
of this factor \\ 
$ N \ (h_{+}+h_{-}+1)^{\min{(l_x,l_y)}}$ 
for the necessary memory 
in calculating the series to order $u^N$ 
by this improved algorithm. 

 It is quite natural to introduce the partition function 
for the finite-size lattice with the range of the variable $h_i$ 
restricted as in equation~(\ref{defH}). 
It comes from the analogy with the algorithm 
of the low-temperature expansion 
for the surface tension in $d=3$ Ising model 
by the finite-lattice method
      \cite{Arisuefour}. 
 We should stress that it enables us to calculate much longer series 
than the case when we would apply the finite-lattice method naively 
by considering the ASOS model as a two-dimensional spin system. 
 In the latter case we would define 
\begin{equation}
   H(\Lambda) 
     = - \ln{ \big[ Z(\Lambda) \ \big] }     
      \label{defHprime}
\end{equation}
and 
\begin{equation}
  W(\Lambda)
    = H(\Lambda) 
     - {\sum_{\scriptstyle \Lambda^{\prime} \subset \Lambda } }
      W( \Lambda^{\prime} ). 
\end{equation}
for the rectangular $l_x \times l_y$ lattice $\Lambda$. 
Here the variable $h_i$ takes all integer values 
$ - \infty \le h_i \le +\infty $ in the calculation 
of the partition function $Z(\Lambda)$. 
 Then the free energy density in the infinite-volume limit 
would be given by 
\begin{equation}
    f =  \sum_{ l_x,l_y} W( l_x \times l_y ).
      \label{free_energy2}      
\end{equation}
 For the expansion series to order $u^N$, 
we should take into accout all the rectangular lattices 
that satisfy 
$ l_x + l_y \le N $
in the summation of equation~(\ref{free_energy2}). 
The necessary CPU time and memory to evaluate each $Z(\Lambda)$ 
to order $u^N$ for $l_x \times \l_y$ lattice 
is proportional to $N l_x l_y (2[N/2]+1)^{\min{(l_x,l_y)}}$ 
and $N (2[N/2]+1)^{\min{(l_x,l_y)}}$, respectively. 
 In table~\ref{tab:memory} we also list the maximum $M_{naive}(N)$ 
of this factor $N (2[N/2]+1)^{\min{(l_x,l_y)}}$ 
for the necessary memory in calculating the series to order $u^N$ 
by this naive algorithm. 
 We can see that the necessary memory in the improved algorithm 
is much smaller than in the naive algorithm.

 We note that the improved algorithm above can be used 
in the finite-lattice method of the low-temperature expansion 
for the spin systems whose spin variable takes more than 2 values 
such as the spin-$S$ Ising model with $S \ge 1$ 
or the $q$-state Potts model with $q \ge 3$.
 
\section{Series}
 We have calculated the expansion series for the free energy $f$
of ASOS model to order $u^{23}$. 
 The calculation was performed on FACOM-VP2600 
at Kyoto University Data Processing Center. 
 Used main memory was about 80Mbyte
and total CPU time was about 25 minutes.
 The obtained coefficients for the free energy are listed 
in table~\ref{tab:series}, 
where $a_n$ is defined as 
\begin{equation}
   f =  \sum_n a_n u^n.                         \label{eqfe}
\end{equation}
 We have checked that each of the $W(l_x \times l_y;h_{+},h_{-})$'s 
in equation~(\ref{free_energy}) starts from the correct order in $u$ 
discussed in the previous section.
 The series to order $u^{12}$ for ASOS model coincides 
with those obtained by Hasenbusch and Pinn
          \cite{Hasenbuschtwo}
and we have obtained new 11 terms.
 Also listed are the coefficients for the surface tension $\sigma$
in $d=3$ simple cubic Ising model obtained previously by the author
          \cite{Arisuefour}.
The coefficients are defined by
\begin{equation}
   \sigma =  2 \beta_{Ising} + \sum_n b_n u_{Ising}^n ,
\end{equation}
where $u_{Ising}=\exp{(-4\beta_{Ising})}$ is 
the low-temperature expansion parameter in the Ising model.

\section{Series Analysis}
 It is to be noticed that the coefficients of the free energy 
in ASOS model are all negative to order $u^{10}$ 
and their sign changes at the order of $u^{11}$ \cite{Hasenbuschtwo}. 
 We find no another change of sign in higher orders to $u^{23}$.
 This property of the sign-change 
is common to the surface tension in $d=3$ Ising model.
 As was pointed out in reference~\cite{Hasenbuschtwo} 
it supports the claim that the roughening transition of these models 
is of Kosterlitz-Thouless type.
 Let us consider a K-T type function and its expansion in $u$ as 
\begin{equation}
     f^{KT}(u) = A u^2 \exp{[-c (u_r-u)^{-1/2}]}
                 =  \sum_n a^{KT}_n u^n                    
\end{equation}
and fit this expansion to the expansion of the free energy 
(\ref{eqfe}) of ASOS model, so that the ratios of 
their successive coefficients coincide 
as $a_{n}/a_{n-1}=a^{KT}_{n}/a^{KT}_{n-1}$ 
for $n=11$ and $n=23$, respectively. 
The order $n=11$ is just where the change of sign occurs
and $n=23$ is the highest order of our series.
 Then we obtain $u_r=0.214$ and $c = 0.533$. 
 In table~\ref{tab:ratio} we list the ratios $a_{n}/a_{n-1}$ 
for the free energy and the fitted K-T type function. 
 We note that the ratios agree with each other within 3 per cent 
for the range of $n = 6$ to $23$. 
%
%
 Furthermore the fitted value of $u_r$ is 
not far from the predicted roughening transition point for ASOS model as
$u_r=0.207(9)$ from the series analysis of the surface width 
        \cite{Adler}
and
$u_r=0.1994(1)$ from the Monte Carlo renormalization-group analysis 
        \cite{Hasenbuschthree}. 
The difference between the ratios for $n \le 5$ would come 
from the regular background term in the free energy, 
whose contribution can be expected to become negligible rapidly 
for higher orders compared with the contribution from the singular term.
%
%

 Next we will try the Pad\'e approximation for the free energy $f(u)$ 
and the internal energy $E(u)=df/d\beta=-2uf'(u)$, 
although the free energy is expected to behave 
like equation~(\ref{KT}) 
and one might think that the Pad\'e approximation would not be 
appropriate for such a function. 
The authors of reference~\cite{Hasenbuschtwo} stressed 
the discrepancy between the Pad\'e approximants 
of the internal energy to order $u^{12}$ and their Monte Carlo data. 
Their series are, however, too short to give a definite conclusion. 

 In figures~1 and 2 we show the truncated series of the free energy 
and its Pad\'e approximants for each order $n$ 
of the series in $\beta=0.85(u=0.1827)$ and $\beta=0.81(u=0.1979)$, 
respectively. 
 The latter point of $\beta$ is very close to the predicted 
roughening transition point
              \cite{Hasenbuschthree}. 
 As for the Pad\'e approximants 
we plot the $[L,M]$ approximants 
($L+M=n-2$) with $ (n-2)/4 \le L \le 3(n-2)/4$. 
Defective approximants 
that have poles on the real axis of $u$ 
in the range of $0 \le u \le 0.205$ are excluded. 
 The Pad\'e approximants appear to converge once to some value 
for the shorter series. 
They begin to scatter just around the order of the sign-change 
and  they  converge again for higher order series but to another value. 
 The convergence is so good in higher orders. 
 The convergence of the truncated series is, on the other hand, 
much slower than the Pad\'e approximants.
 In figure~3 we also give the behaviour 
of the Pad\'e approximants 
together with the truncated series of the free energy in $\beta$ 
for the highest order series. From 
these figures we can conclude that the Pad\'e approximation 
works well for the long enough series of the free energy of ASOS model 
even very close to the predicted roughening transition point. 
 This situation of the Pad\'e approximants 
for the free energy of ASOS model 
is quite similar to the case for the surface tension 
in $d=3$ Ising model  
      \cite{Shaw,Arisuefour},
in which presently available series has only 5 terms 
after the sign-change. 
 We can expect that 
the higher order series for the surface tension in $d=3$ Ising model 
will also give Pad\'e approximants converging to the Monte Carlo data 
and the present discrepancy \cite{Hasenbuschtwo} will be resolved. 

 In figures~4 and 5 we show the truncated series 
of the internal energy and its Pad\'e approximants 
for each order $n$ of the series 
for $\beta=0.85$ and $\beta=0.81$, respectively. 
 We see that the behaviors of the truncated series 
and its Pad\'e approximants in the order $n$ 
are similar to those for the free energy given in figures~1 and 2. 
 The convergence of the Pad\'e approximants is good 
for higher orders, 
although the convergence for $\beta=0.81$ 
is slower than in the case of $\beta=0.85$. 
 On the contrary, 
the convergence of the truncated series is much slower 
than the Pad\'e approximants. 
 In figure~6 and table~\ref{tab:energy} we also give the behaviour of 
the Pad\'e approximants for the internal energy versus $\beta$ 
for the highest order series. 
The Pad\'e approximants for the highest order series are consistent 
with the Monte Carlo data 
      \cite{Hasenbuschtwo}
even at $\beta=0.81$ very close to the roughening transition point.



 
 One's question might be why the naive Pad\'e approximation,
which is the ratio of two polynomials, 
is so good even close to the roughening transition point 
in spite of that the free energy is 
expected to have essential singularity as in equation~(\ref{KT}). 
To clarify the reason we examine the Pad\'e approximation for 
a test function
\begin{equation}
     g(x) =  \exp{[- (1-x)^{-1/2}]} + \exp{(x)}.
\end{equation} 
 The first term has the essential singularity at $x=1$ 
and the second term is a regular background. 
 In figures~7 and 8 we plot the exact function, 
the 23rd order truncated series and its Pad\'e approximants 
of the test function $g(x)$ 
and of its derivative $x g^{'}(x)$, respectively. 
 As for the test function itself, 
we can see that the Pad\'e approximants exhibit so good convergence 
and give the values that are consistent with the exact function 
for $x \le 0.98$. 
 As for the derivative, 
we observe that the Pad\'e approximants converge 
and follow the exact function well for $x \leq 0.95$. 
 The derivative shows worse convergence than the test function itself, 
which would be because the derivative has 
the power singularity multiplied by the essential singularity. 
 The truncated series, on the contrary, starts to deviate 
from the exact function much earlier than the Pad\'e approximants. 
 We also plot in figures~9 and 10 the exact value, truncated series 
and its Pad\'e approximants 
of the free energy and the internal energy, respectively, 
for each order $n$ of the series in $x=0.98$. 
 The situation of the test function in figures~7 and  9 
is quite similar to that of the free energy of ASOS model 
in figures~3 and 2 
and the situation of the derivative in figures~8 and 10 
is quite similar to 
the internal energy of ASOS model in figures~6 and 5, respectively. 
 We have examined several other test functions that have the same 
structure as equation~(\ref{KT}) and the situations are 
all similar to the case of the test function $g(x)$. 
 We can conclude that 
the Pad\'e approximants of the function like equation~(\ref{KT}), 
which involves an essential singularity, converges 
to the correct value even very close to the singularity point, 
if we use the expansion series in high enough order. 

\section{Summary}

 We have calculated the low-temperature series 
for the free energy of ASOS model to order $u^{23}$ 
using the finite-lattice method. 
 An improved technique has been used to extend the series, 
which can also be applied to the low-temperature expansion 
of the spin systems whose spin-variable takes more than 2 values. 
 The property of the obtained series 
and the behavior of their Pad\'e approximants have confirmed us 
that the roughening transition in the ASOS model is of 
Kosterlitz-Thouless type. 

\section*{Acknowledgements}
\hspace*{\parindent}
 The author would like to thank 
K. Tabata, K. Pinn, M. Hasenbusch, G. M\"unster and P. Weisz 
for valuable discussions. 
%

\begin{table}[htb]
\caption{
Comparison of the factors $M_{imp}(N)$ for the improved algorithm 
and $M_{naive}(N)$ for the naive algorithm 
which the necessary memory is proportional to 
in calculating the series to order $u^N$. 
         }
\label{tab:memory}
\begin{center}
\begin{tabular}{|r|r|r|}
\hline
   $N$  & $M_{imp}(N)$ & $M_{naive}(N) \ \ \ \ \ $ \\
\hline
$ 2$  & $       4$  &   $                  6$ \\
$ 3$  & $       6$  &   $                  9$ \\
$ 4$  & $      16$  &   $                100$ \\
$ 5$  & $      20$  &   $                125$ \\
$ 6$  & $      54$  &   $               2058$ \\
$ 7$  & $      63$  &   $               2401$ \\
$ 8$  & $     216$  &   $              52488$ \\
$ 9$  & $     243$  &   $              59049$ \\
$10$  & $     810$  &   $            1610510$ \\
$11$  & $     891$  &   $            1771561$ \\
$12$  & $    3072$  &   $           57921708$ \\
$13$  & $    3328$  &   $           62748517$ \\
$14$  & $   14336$  &   $         2392031250$ \\
$15$  & $   15360$  &   $         2562890625$ \\
$16$  & $   65536$  &   $       111612119056$ \\
$17$  & $   69632$  &   $       118587876497$ \\
$18$  & $  294912$  &   $      5808378560022$ \\
$19$  & $  311296$  &   $      6131066257801$ \\
$20$  & $ 1562500$  &   $    333597619564020$ \\
$21$  & $ 1640625$  &   $    350277500542221$ \\
$22$  & $ 8593750$  &   $  20961814674106394$ \\
$23$  & $ 8984375$  &   $  21914624432020321$ \\
$24$  & $46875000$  &   $1430511474609375000$ \\
$25$  & $48828125$  &   $1490116119384765625$ \\
\hline
\end{tabular}
\end{center}

\end{table} 
\begin{table}[htb]
\caption{
The low-temperature expansion coefficients $a_n$ 
for the free energy of ASOS model
and $b_n$ for the surface tension of 3-d Ising model.
         }
\label{tab:series}
\begin{center}
\begin{tabular}{|r|l|l|}
\hline
   $n$  & \multicolumn{1}{c|}{$a_n$} & \multicolumn{1}{c|}{$b_n$} \\
\hline
$ 2$  & $-            2   $  &   $- 2            $ \\
$ 3$  & $-            4   $  &   $- 2            $ \\
$ 4$  & $-           10   $  &   $- 10           $ \\
$ 5$  & $-           24   $  &   $- 16           $ \\
$ 6$  & $-          194/ 3$  &   $- 242 / 3      $ \\
$ 7$  & $-          172   $  &   $- 150          $ \\
$ 8$  & $-          452   $  &   $- 734          $ \\
$ 9$  & $-         3184/ 3$  &   $- 4334 / 3     $ \\
$10$  & $-         8862/ 5$  &   $- 32122 / 5    $ \\
$11$  & $+         1712   $  &   $- 10224        $ \\
$12$  & $+       118604/ 3$  &   $- 106348 / 3   $ \\
$13$  & $+       284232   $  &   $+ 53076        $ \\
$14$  & $+     11450234/ 7$  &   $+ 3491304 / 7  $ \\
$15$  & $+     42775116/ 5$  &   $+ 74013814 / 15$ \\
$16$  & $+     42439230   $  &   $+ 27330236     $ \\
$17$  & $+    203605872   $  &   $+ 160071418    $ \\
$18$  & $+   8589601858/ 9$  &   $               $ \\
$19$  & $+   4397057716   $  &   $               $ \\
$20$  & $+  19982814764   $  &   $               $ \\
$21$  & $+1885469273984/21$  &   $               $ \\
$22$  & $+4393383644538/11$  &   $               $ \\
$23$  & $+1760491870744   $  &   $               $ \\
\hline
\end{tabular}
\end{center}
\end{table} 

\begin{table}[htb]
\caption{
Comparison of the ratio of ${a_{n}}/{a_{n-1}}$ 
for the free energy 
and ${a^{KT}_{n}}/{a^{KT}_{n-1}}$ 
for the fitted K-T type function $f^{KT}(u)$. 
The parameter for the K-T type function is 
$u_r=0.214$ and $c=0.533$.
         }
\label{tab:ratio}
\begin{center}
\begin{tabular}{|r|r|r|}
\hline
   $n$  & ${a_{n}}/{a_{n-1}}$ & ${a^{KT}_{n}}/{a^{KT}_{n-1}}$ \\
\hline
$ 3$  & $ 2.000$  &   $-2.692$ \\
$ 4$  & $ 2.500$  &   $ 2.158$ \\
$ 5$  & $ 2.400$  &   $ 2.510$ \\
$ 6$  & $ 2.694$  &   $ 2.658$ \\
$ 7$  & $ 2.660$  &   $ 2.683$ \\
$ 8$  & $ 2.628$  &   $ 2.591$ \\
$ 9$  & $ 2.348$  &   $ 2.321$ \\
$10$  & $ 1.670$  &   $ 1.634$ \\
$11$  & $-0.966$  &   $-0.966$ \\
$12$  & $23.093$  &   $22.345$ \\
$13$  & $ 7.189$  &   $ 7.029$ \\
$14$  & $ 5.755$  &   $ 5.635$ \\
$15$  & $ 5.230$  &   $ 5.126$ \\
$16$  & $ 4.961$  &   $ 4.869$ \\
$17$  & $ 4.798$  &   $ 4.717$ \\
$18$  & $ 4.687$  &   $ 4.618$ \\
$19$  & $ 4.607$  &   $ 4.549$ \\
$20$  & $ 4.545$  &   $ 4.499$ \\
$21$  & $ 4.493$  &   $ 4.461$ \\
$22$  & $ 4.448$  &   $ 4.432$ \\
$23$  & $ 4.408$  &   $ 4.408$ \\
\hline
\end{tabular}
\end{center}

\end{table} 

\begin{table}[htb]
\caption{
Comparison of the Pad\'e approximants of the low-temperature series 
and the Monte Carlo data in reference~[15] 
for internal energy. 
The value of the truncated series is also listed.
         }
\label{tab:energy}
\begin{center}
\begin{tabular}{|l|l|l|l|l|l|}
\hline                                                                
\multicolumn{1}{|c|}{$\beta$} &  $0.85$  &  $0.84$  &  $0.83$  &   $0.82$ &  $0.81$  \\
\hline                                                                
\multicolumn{1}{|c|}{$u$} & $0.1827$ & $0.1864$ & $0.1901$ & $0.1940$ &  $0.1979$\\
\hline                                                                
\hline                                                                
\multicolumn{1}{|c|}{truncated}   & $0.5917$ & $0.6252$ & $0.6590$ & $0.6919$ & $0.7222$ \\
\hline                                                                
Pad\'e \hfill $[6/15] $ \  & $0.5886$ & $0.6202$ & $0.6509$ & $0.6797$ & $0.7060$ \\
\hfill $[7/14] $ \   & $0.5886$ & $0.6201$ & $0.6508$ & $0.6792$ & $0.7045$ \\
\hfill $[8/13] $ \   & $0.5886$ & $0.6202$ & $0.6509$ & $0.6795$ & $0.7054$ \\
\hfill $[9/12] $ \   & $0.5887$ & $0.6203$ & $0.6511$ & $0.6803$ & $0.7081$ \\
\hfill $[10/11]$ \   & $0.5887$ & $0.6203$ & $0.6511$ & $0.6803$ & $0.7079$ \\
\hfill $[11/10]$ \   & $0.5887$ & $0.6203$ & $0.6512$ & $0.6806$ & $0.7089$ \\
\hfill $[12/9] $ \   & $0.5887$ & $0.6203$ & $0.6513$ & $0.6807$ & $0.7094$ \\
\hfill $[13/8] $ \   & $0.5886$ & $0.6202$ & $0.6509$ & $0.6797$ & $0.7059$ \\
\hfill $[14/7] $ \   & $0.5886$ & $0.6202$ & $0.6509$ & $0.6796$ & $0.7059$ \\
\hfill $[15/6] $ \   & $0.5886$ & $0.6201$ & $0.6507$ & $0.6790$ & $0.7039$ \\
\hline                                                                
MC($L=256$) & $0.5887(2)$ & $0.6205(2)$ & $0.6518(2)$ & $0.6811(2)$ & $0.7096(2)$ \\
MC($L=128$) & $0.5884(4)$ & $0.6203(4)$ & $0.6516(4)$ & $0.6805(4)$ & $0.7095(3)$ \\
MC($L=\ 64$)  & $0.5874(8)$ & $0.6498(7)$ & $0.6498(7)$ & $0.6795(6)$ & $0.7074(6)$ \\
\hline
\end{tabular}
\end{center}

\end{table} 
\clearpage
\newpage
\pagestyle{empty}
\section*{Figure captions}
\hspace*{\parindent}
\parskip=0.5cm

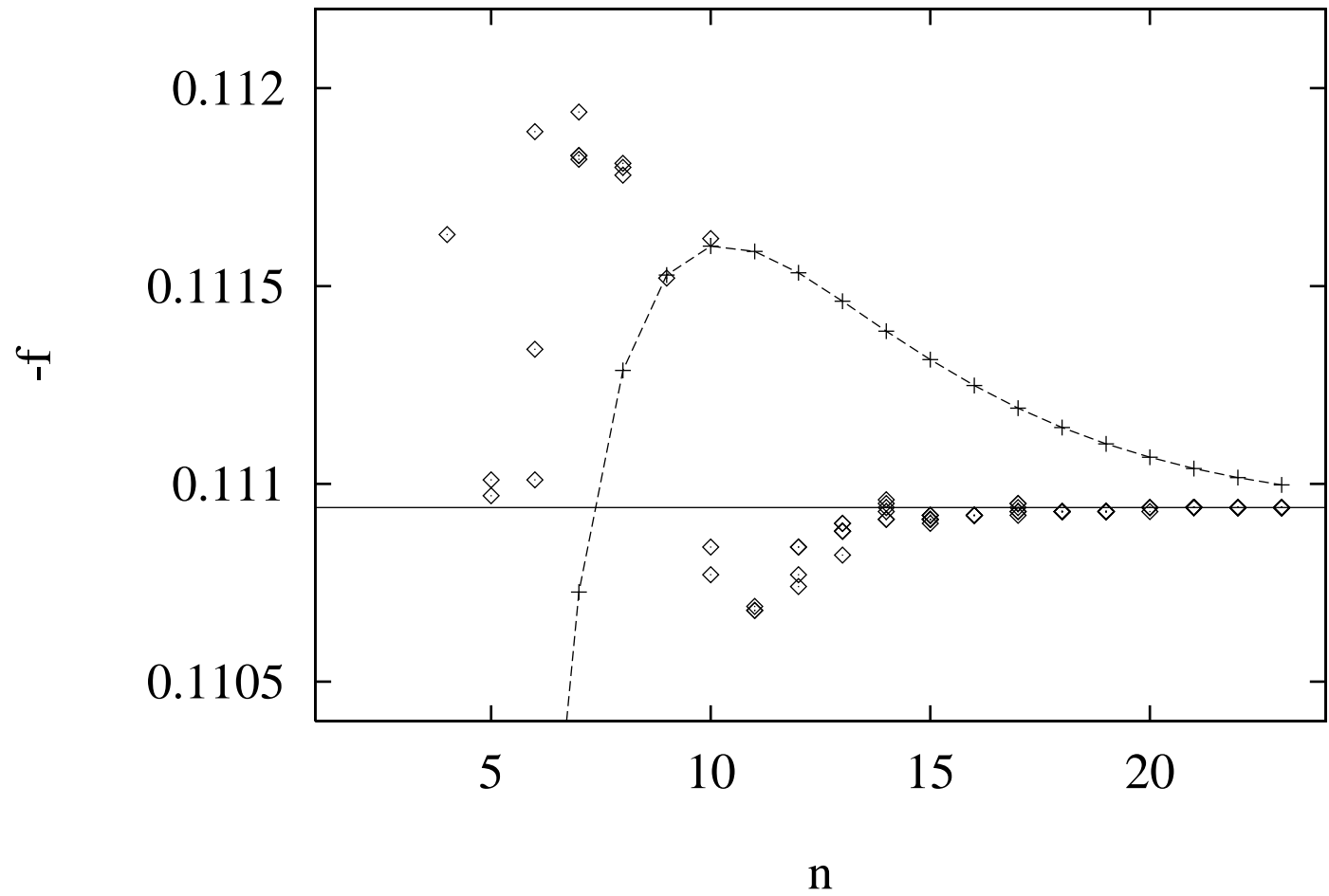
Fig. 1 : 
The values of the truncated series ($+$)
and its Pad\'e approximants ($\Diamond$) for the free energy 
versus the order $n$ of the series 
in $\beta=0.85(u=0.1827)$. 
 The horizontal line represents the average of the Pad\'e approximants 
for the highest order series of $n=23$.

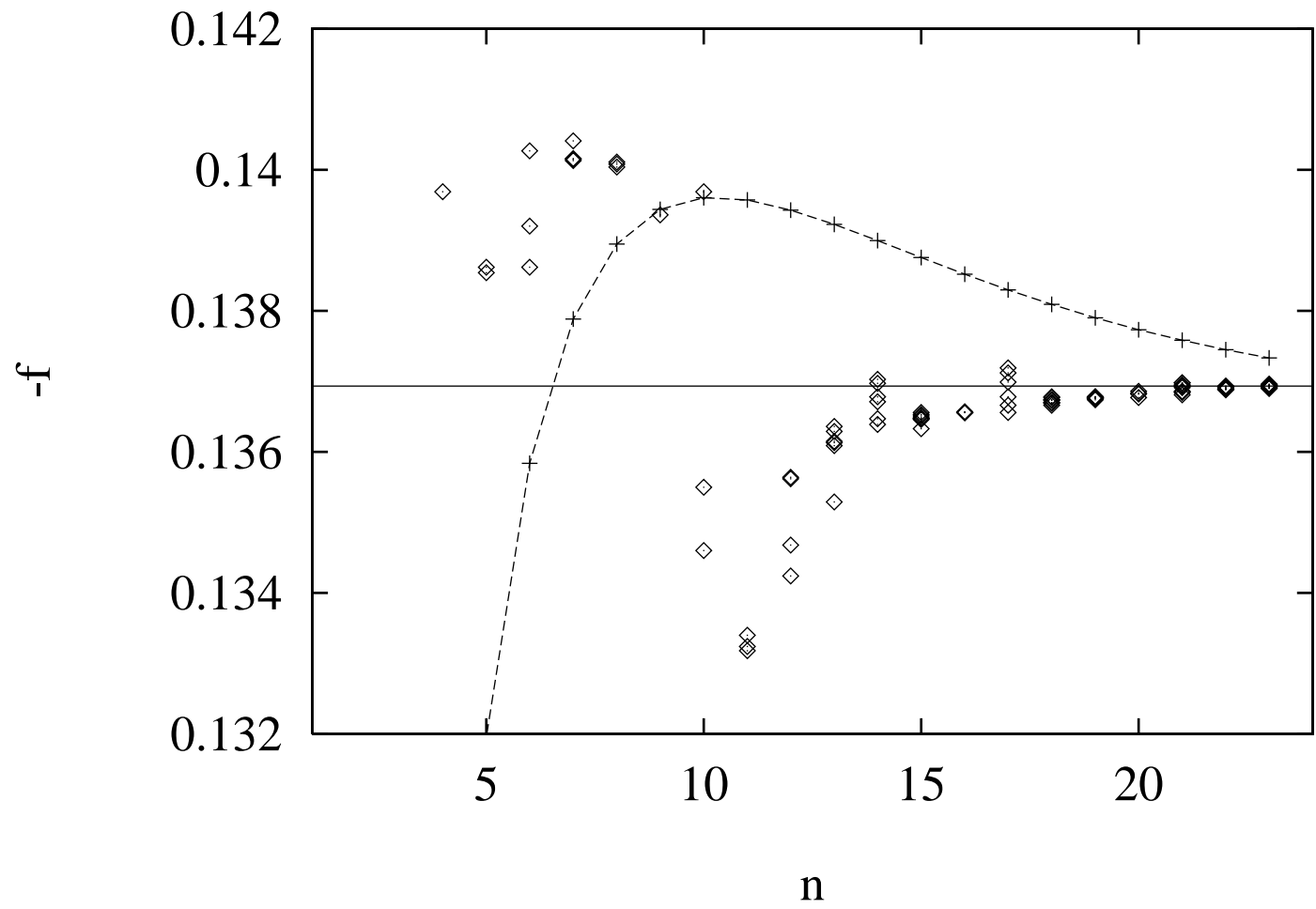
Fig. 2 :
The values of the truncated series ($+$)
and its Pad\'e approximants ($\Diamond$) for the free energy 
versus the order $n$ of the series 
in $\beta=0.81(u=0.1979)$.
 The horizontal line represents the average of the Pad\'e approximants 
for the highest order series of $n=23$.

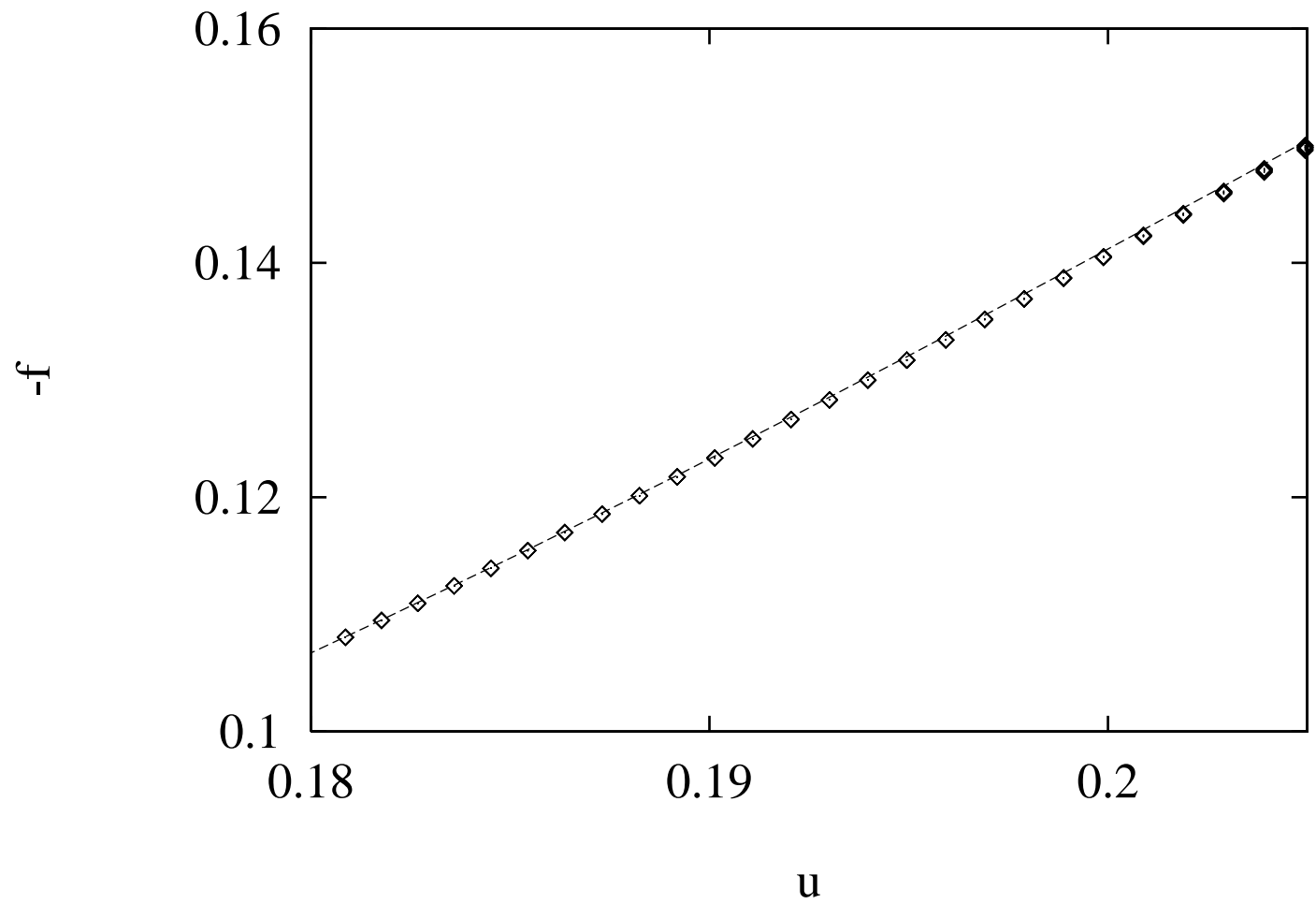
Fig. 3 :
The Pad\'e approximants ($\Diamond$)
together with the truncated series (dashed line) of the free energy 
versus $u$ 
for the highest order series of $n=23$.

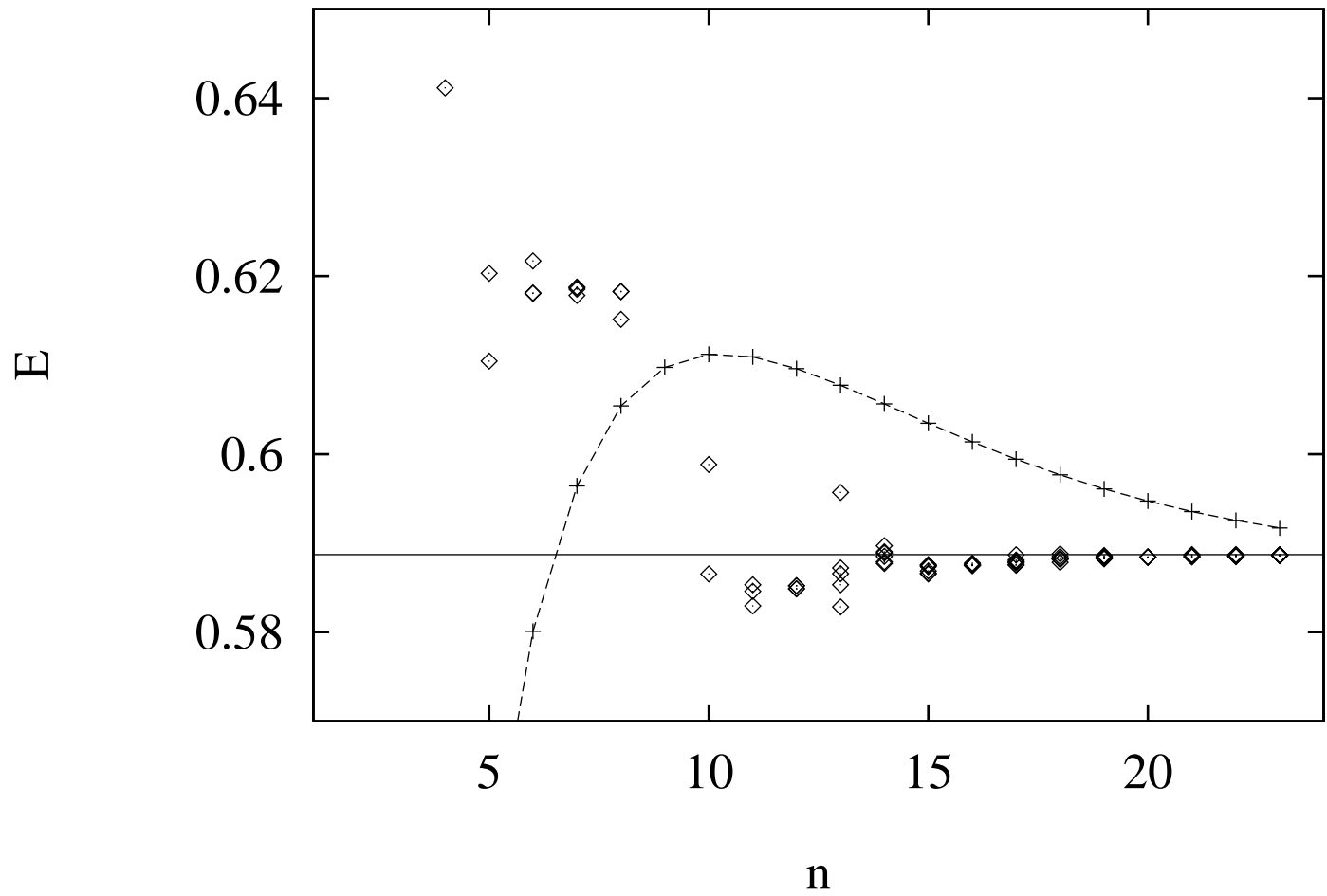
Fig. 4 :
The values of the truncated series ($+$)
and its Pad\'e approximants ($\Diamond$) for the internal energy 
versus the order $n$ of the series 
in $\beta=0.85$.
 The horizontal line represents the result 
of the Monte Carlo simulation in reference~\cite{Hasenbuschtwo}.

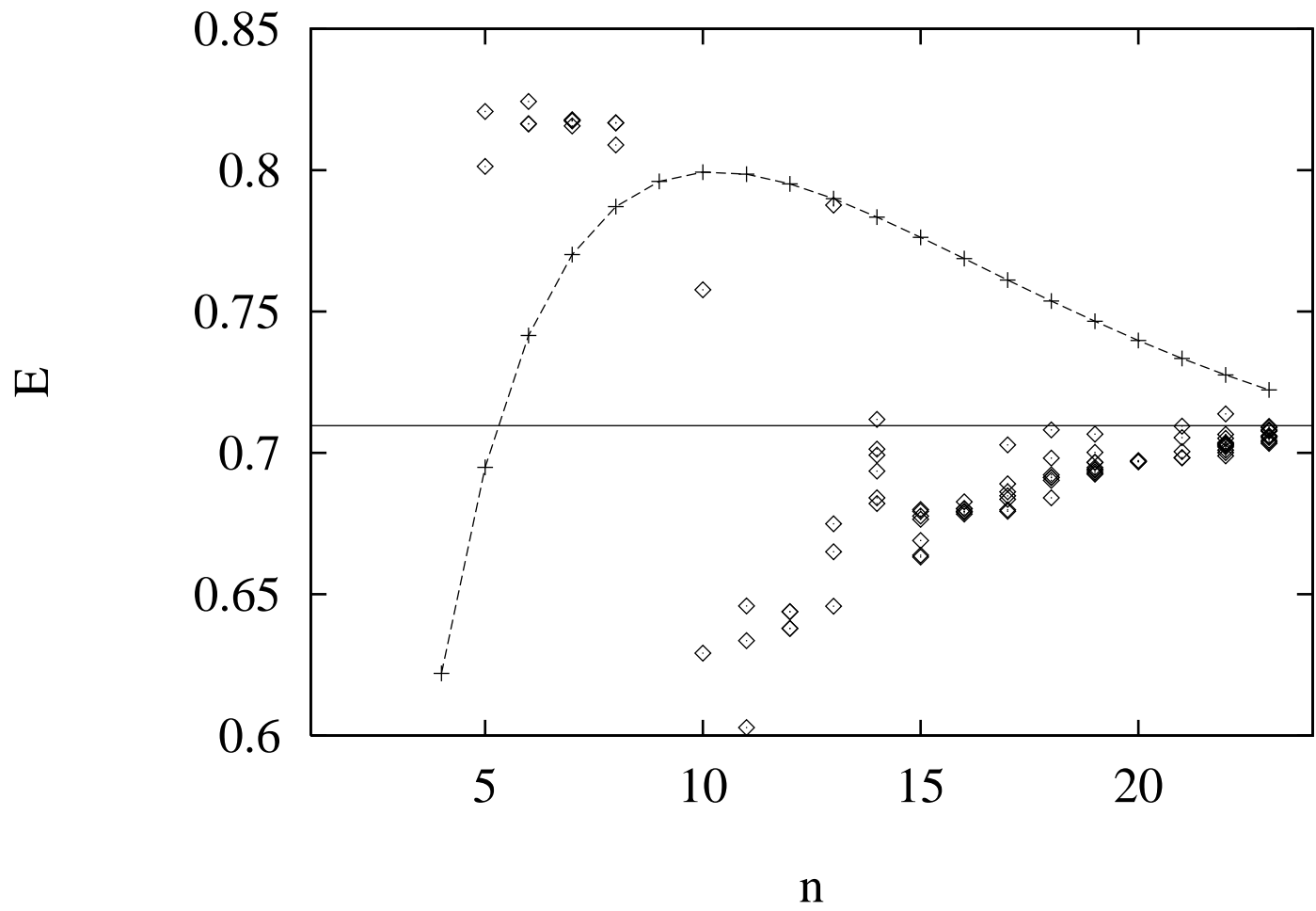
Fig. 5 :
The values of the truncated series ($+$)
and its Pad\'e approximants ($\Diamond$) for the internal energy 
versus the order $n$ of the series 
in $\beta=0.81$.
 The horizontal line represents the result 
of the Monte Carlo simulation in reference~\cite{Hasenbuschtwo}.

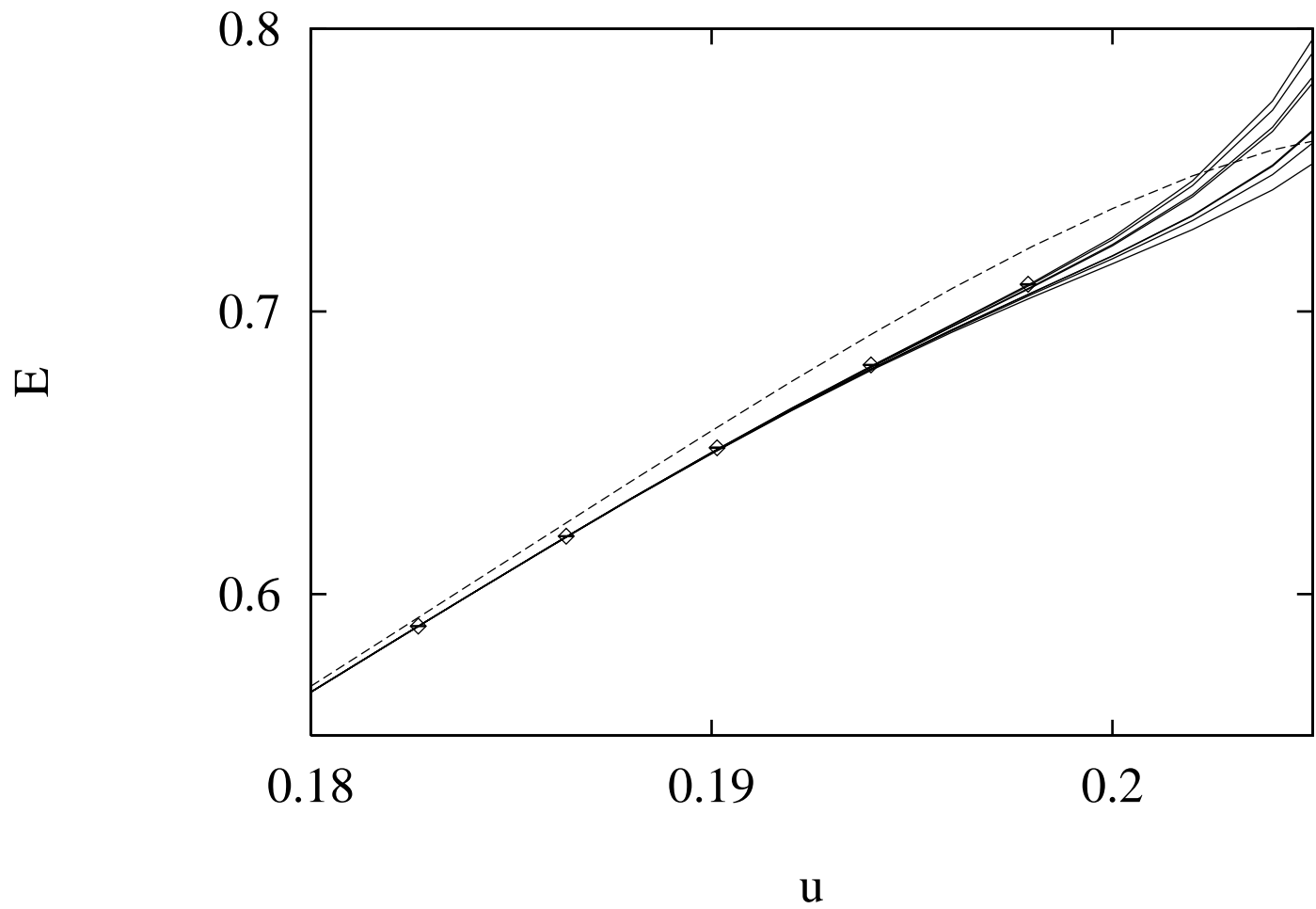
Fig. 6 :
The Pad\'e approximants (solid lines)
together with the truncated series (dashed line) of the internal energy 
versus $u$ 
for the highest order series of $n=23$.
The result of the Monte Carlo simulation ($\Diamond$) 
\hspace{-0.77cm} \raisebox{.04cm}{$-$} 
in reference~\cite{Hasenbuschtwo} is also plotted.

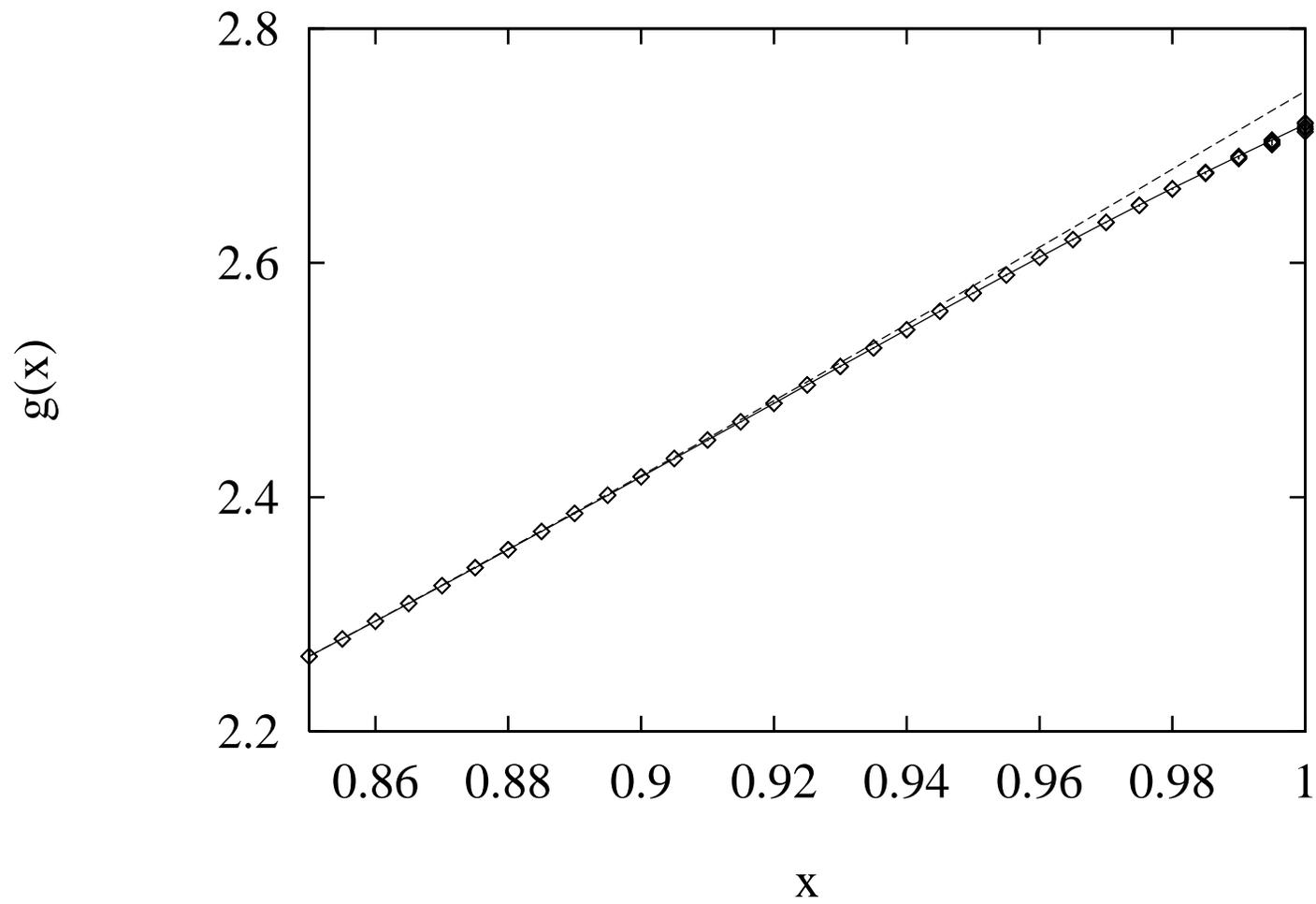
Fig. 7 :
The exact function (solid line), 
the 23rd order truncated series (dashed line) 
and its Pad\'e approximants ($\Diamond$)
of the test function $g(x)$. 

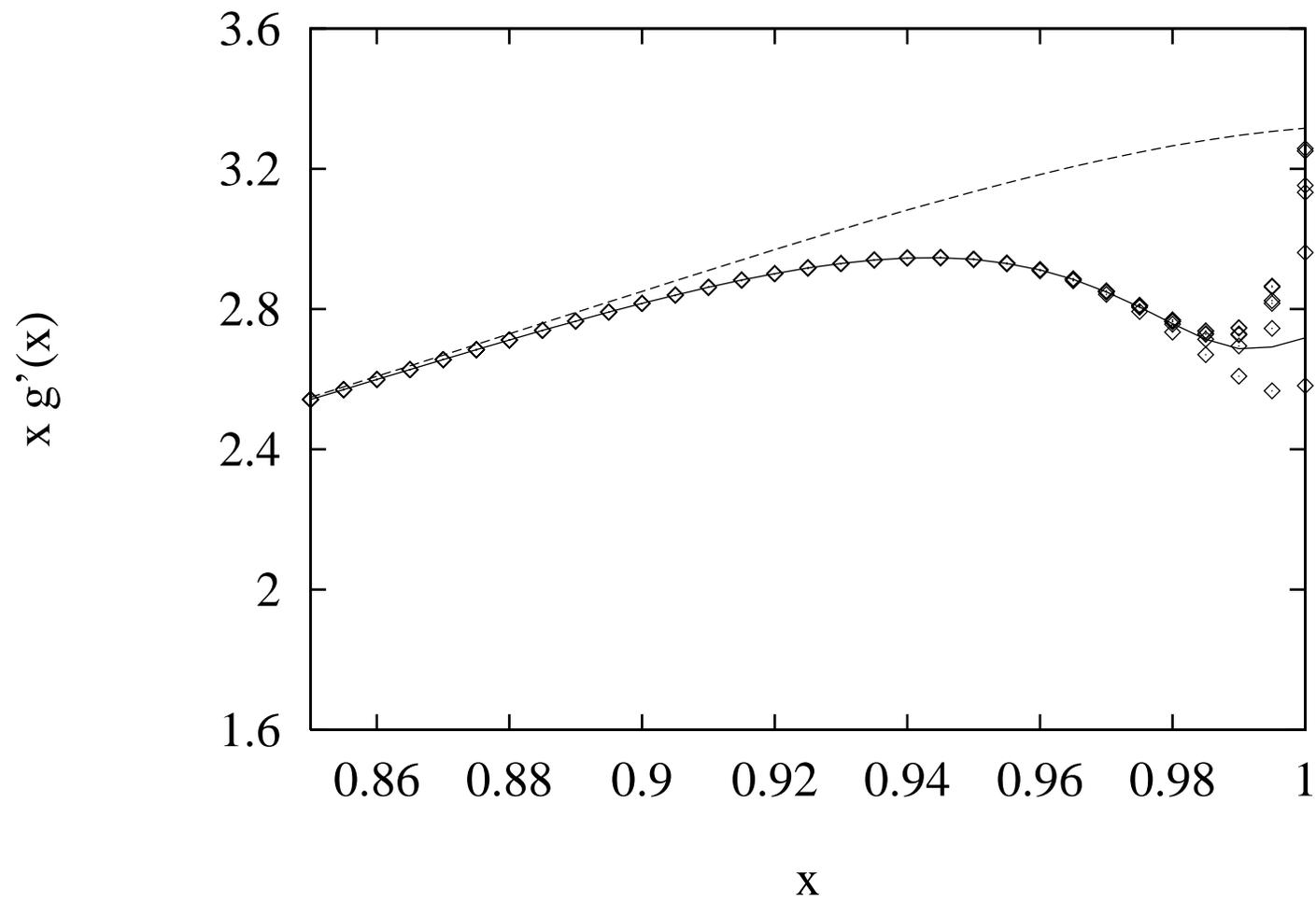
Fig. 8 :
The exact function (solid line), 
the 23rd order truncated series (dashed line) 
and its Pad\'e approximants ($\Diamond$)
of the derivative $x g^{'}(x)$ of the test function. 

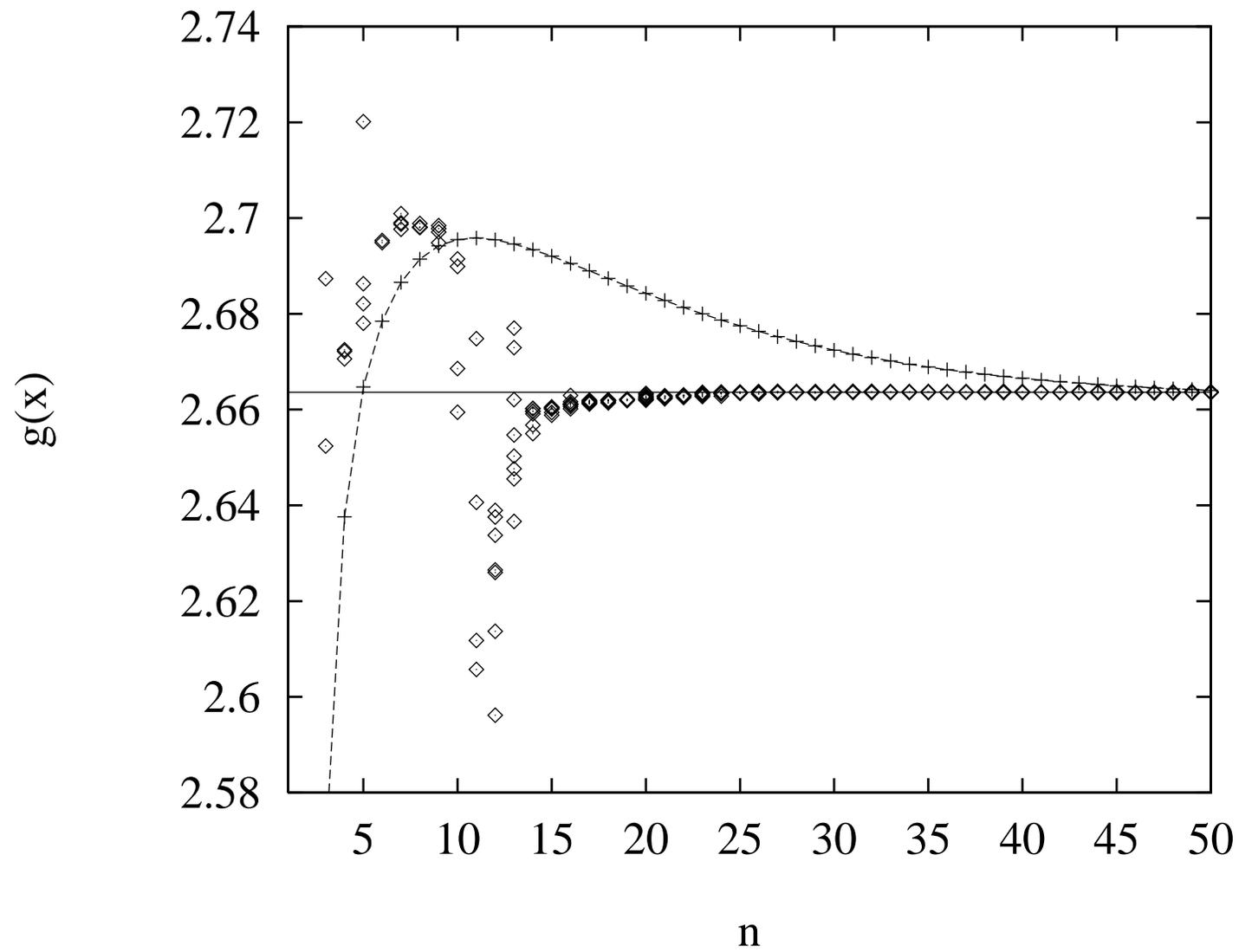
Fig. 9 :
The values of the truncated series ($+$)
and its Pad\'e approximants ($\Diamond$)
for the test function $g(x)$
versus the order $n$ of the series 
in $x=0.98$.
 The horizontal line represents the exact value.

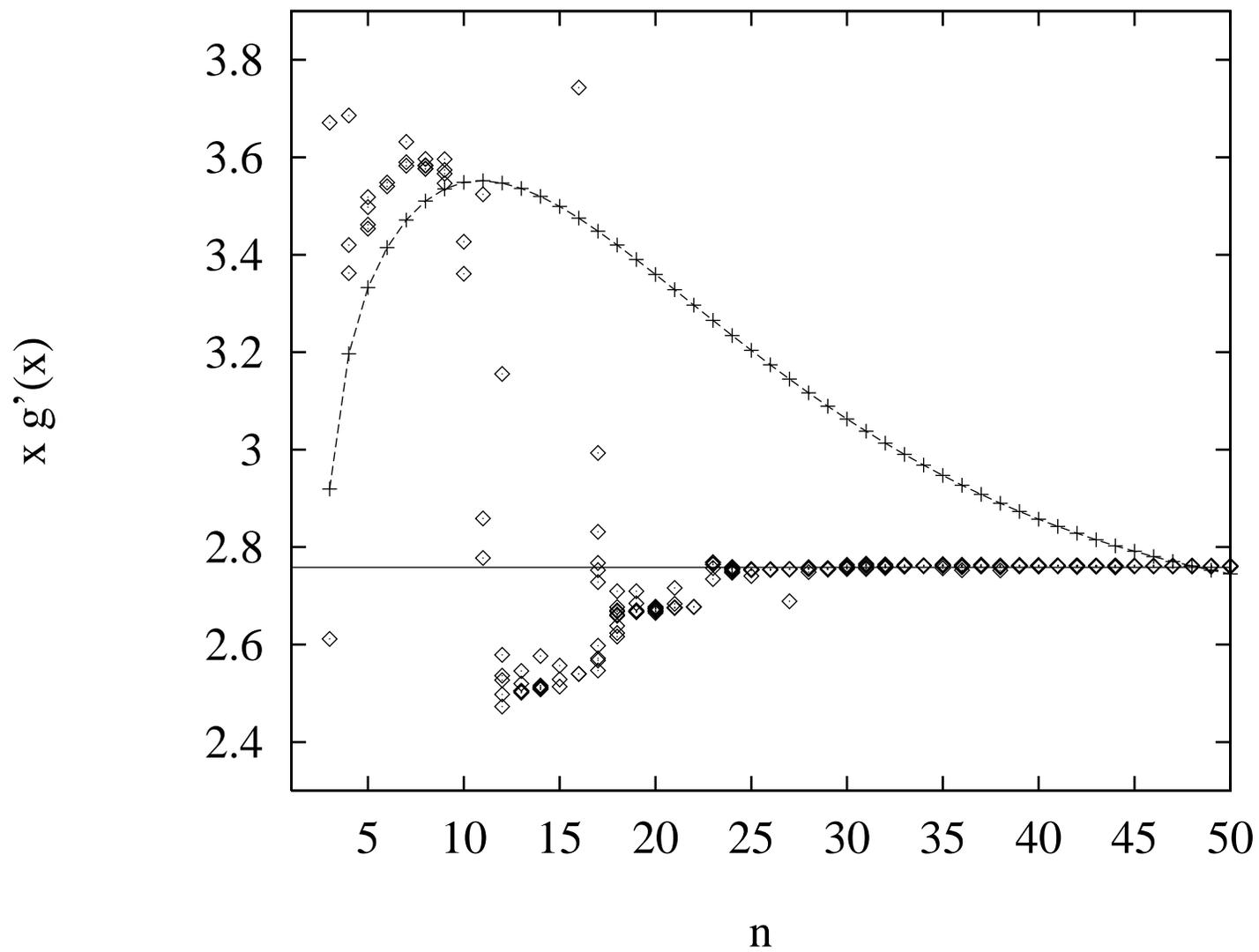
Fig. 10 :
The values of the truncated series ($+$)
and its Pad\'e approximants ($\Diamond$)
for the derivative $xg^{'}(x)$ of the test function 
versus the order $n$ of the series 
in $x=0.98$.
 The horizontal line represents the exact value.

\end{document}